\documentclass[runningheads]{llncs}
\usepackage{graphicx}
\usepackage{enumitem}
\usepackage{url}
\usepackage[linesnumbered,ruled,vlined]{algorithm2e}
\usepackage{caption}
\usepackage{orcidlink}
\usepackage{subcaption}

\begin{document}
\title{Data-Driven Intelligence can Revolutionize Today's Cybersecurity World: A Position Paper}
\titlerunning{Data-Driven Intelligence can Revolutionize Today's Cybersecurity World}
%
\author{Iqbal H. Sarker\textsuperscript{1,2,*}\orcidlink{0000-0003-1740-5517}, Helge Janicke\textsuperscript{1,2}\orcidlink{0000-0002-1345-2829}, Leandros Maglaras\textsuperscript{3}\orcidlink{0000-0001-5360-9782} and
Seyit Camtepe\textsuperscript{4}\orcidlink{0000-0001-6353-8359}}
\authorrunning{Sarker et al.}
%
\institute{\textsuperscript{1} Security Research Institute, Edith Cowan University, Perth, WA-6027, Australia.\\ \textsuperscript{2} Cyber Security Cooperative Research Centre, Australia. \\ \textsuperscript{3} School of Computing, Edinburgh Napier University, Edinburgh EH10 5DT, UK. \\  \textsuperscript{4} Data61, CSIRO, Sydney, NSW 2122, Australia. \\
$^*$Correspondence: m.sarker@ecu.edu.au
}
\maketitle              
\begin{abstract}
As cyber threats evolve and grow progressively more sophisticated, cyber security is becoming a more significant concern in today's digital era. Traditional security measures tend to be insufficient to defend against these persistent and dynamic threats because they are mainly intuitional. One of the most promising ways to handle this ongoing problem is utilizing the potential of \textit{data-driven intelligence}, by leveraging AI and machine learning techniques. It can improve operational efficiency and saves response times by automating repetitive operations, enabling real-time threat detection, and facilitating incident response. In addition, it augments human expertise with insightful information, predictive analytics, and enhanced decision-making, enabling them to better understand and address evolving problems. Thus, data-driven intelligence could significantly improve real-world cybersecurity solutions in a wide range of application areas like critical infrastructure, smart cities, digital twin, industrial control systems and so on. In this position paper, we argue that data-driven intelligence can revolutionize the realm of cybersecurity, offering not only large-scale task \textit{automation} but also \textit{assist human experts} for better situation awareness and decision-making in real-world scenarios.

\end{abstract}
\section{Introduction}
Cybersecurity has emerged as a major problem in today's hyperconnected world due to the growing cyber threat landscape and the increasing number of sophisticated malicious actors. According to the Telecommunication Standardization Sector of International
Telecommunication Union \cite{itu2009overview} ``Cybersecurity is the collection of tools, policies, security
concepts, safeguards, guidelines, risk management approaches, actions, training, best practices, assurance and
technologies that can be used to protect the cyber environment and organization and user’s assets." In the real-world scenario, protecting sensitive data and digital assets from continuously evolving threats is a challenging task for businesses in a variety of application areas such as critical infrastructures, smart city applications, information and operational technology networks, etc. Traditional security solutions might not be sufficient to provide defense against today's persistent and constantly evolving threats in these areas. There is an urgent need for innovative approaches that can effectively counteract the dynamic nature of cyber threats. Therefore, in this paper, we focus on data-driven intelligence, which offers a powerful combination of automation and human assistance and could be one of the most promising strategies for solving this ongoing problem. 

Data-driven intelligence typically can be defined as the process of using data analysis and interpretation to derive insights or useful knowledge, and eventually make intelligent decisions. It thus involves identifying trends, patterns, correlations, and other pertinent information primarily through the use of data, which could then be applied to regulate corporate operations and strategic decisions. The development of data-driven intelligence, powered by machine learning and artificial intelligence \cite{sarker2022multi}, has tremendous potential for revolutionizing cybersecurity in various application areas, discussed briefly in Section \ref{Real-World Application Areas}. Data-driven intelligence has the capability to reveal hidden patterns, detect anomalies and predict potential cyberattacks by utilizing the enormous amounts of data generated from numerous sources, such as network logs, system events, and user behavior. This enables the development of proactive and adaptive defense systems rather than simply relying on predefined rules and signatures, enhancing an organization's capacity to recognize, respond to, and mitigate cyber threats. In addition to automating tasks, cyber analysts can gain deeper insights into the tactics, techniques, and procedures employed by cyber adversaries through the extracted insights from data, discussed briefly in Section \ref{Data Insights and Modeling}.

In order to better understand the main focus of this position paper and overall contributions, we formulate three major questions below:

\begin{itemize}
    \item Can data-driven intelligence
    \textit{automate} the large-scale complex tasks in the context of cybersecurity?
    
    \item Does data-driven intelligence have the potential to \textit{augment} human expertise or knowledge through in-depth understanding as well as to \textit{assist} them in their decision-making process in real-world scenarios?

    \item Is it worthwhile to \textit{rethink} the present cyberspace across a variety of application areas while taking into account the power of data-driven intelligence, particularly in terms of automation and assisting human experts in the domain?
\end{itemize}

Answering these questions, we believe that data-driven intelligence can revolutionize today's cybersecurity world. Towards this, we provide a clear understanding of the potential of data-driven intelligence as well as their applicability and impact from the perspective of next-generation cybersecurity solutions in the following sections. Thus this paper contributes to the ongoing discussion about the role of data-driven modeling and the importance of ensuring that innovative methods are developed and deployed in a manner that maximizes its benefits while minimizing its risks. The ultimate purpose of this paper is not only to highlight data-driven intelligence but also to use the extracted insights or useful knowledge gained from data to make intelligent decisions that improve the current cybersecurity landscape.  

The rest of the paper is organized as follows: Section \ref{Why Data-Driven Intelligence for Cybersecurity Solutions} highlights the significance of data intelligence considering both automating tasks and human experts' decision-making. We discuss data-driven modeling in Section \ref{Data Insights and Modeling}. We also explore the potentiality of data-driven intelligence in various real-world application domains in Section \ref{Real-World Application Areas}. The key challenges and issues are highlighted in Section \ref{Challenges and Research Direction} and finally, Section \ref{Conclusion} concludes this paper.

\section{Why Data-Driven Intelligence for Next-Generation Cybersecurity?}
\label{Why Data-Driven Intelligence for Cybersecurity Solutions}
In the area of cybersecurity, data-driven intelligence offers a substantial contribution to \textit{automation} as well as \textit{assisting human expert decision-making} to solve real-world problems. Human experts may not have the scalability and speed of automated systems, but they do have the capability for critical thought, intuition, and the ability to realize bigger organizational goals as well as ethical concerns when making decisions. The symbiotic relationship between automation and human expertise enables businesses to develop strong cyber defense capabilities, react to threats promptly, and maintain a competitive advantage in the continually evolving landscape of cybersecurity concerns. In this section, we discuss how data-driven intelligence can serve as a strength factor in cybersecurity by automating repetitive processes, anticipating threats, as well as augments human expertise providing useful information.

\begin{figure}[htbp!]
    \centering
    \includegraphics[width=.95\linewidth]{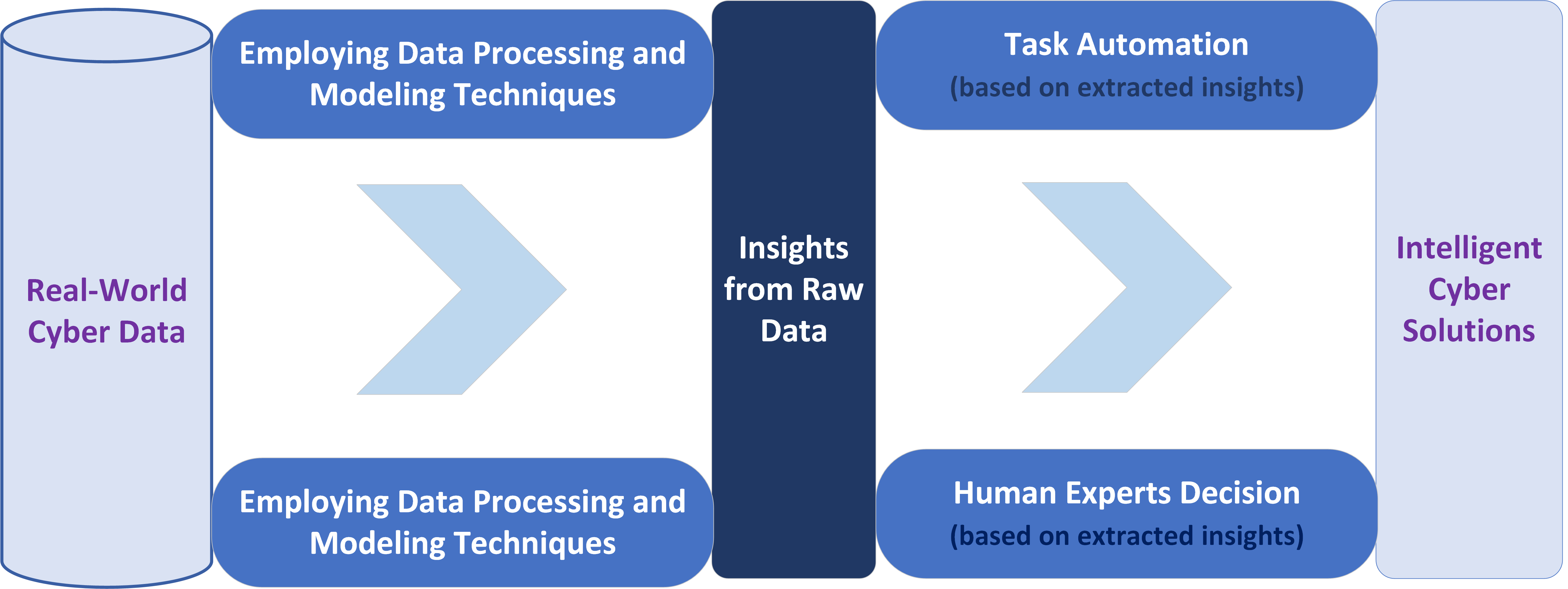}
    \caption{An illustration highlighting the potential of data-driven intelligence for both automation and assisting human experts in the context of cybersecurity.}
    \label{fig: data-driven-intelligence}
\end{figure}

\begin{enumerate}[label=\roman*.]
    \item \textit{Automation of Large-Scale Cyber Tasks:}
    Cybersecurity tasks like log analysis, anomaly detection, and routine security checks can be automated using data-driven intelligence \cite{sarker2021cyberlearning}. Data-driven automated systems use insights from raw data to drive decision-making. These tasks can be completed more quickly and accurately by utilizing machine learning and AI algorithms \cite{sarker2022multi}, alleviating stress on human experts for complicated tasks. By continuously monitoring and analyzing enormous volumes of data from many sources, data-driven intelligence automates the process of threat detection. It instantly detects anomalies, suspicious activity, and potential threats in real time using machine learning techniques. The incident response process is sped up by automation, which also minimizes the risk of human error and ensures that the cybersecurity teams are acting systematically. Through the extraction of insights from raw data, data-driven automated systems are able to continually learn, adapt, and make decisions in real-time, and eventually boost operational effectiveness.

    \item \textit{Augmenting Human Understanding and Expertise for Improved Cyber Solutions:}
    The capabilities of human cybersecurity experts are strengthened by data-driven intelligence in various ways, discussed below. These are - 
        \begin{itemize}
            \item \textit{Assisting Human Experts Decision-Making with Evidence-based Recommendations:}
            Instead of depending exclusively on intuition or prior experiences, cybersecurity professionals could establish complete cybersecurity plans based on empirical evidence and data-informed recommendations with the advancement of data-driven insights. This data-driven approach allows them to conduct comprehensive risk assessments, understand the impact of different attack vectors, and identify critical areas for policy improvement. By providing context-sensitive information about particular incidents and attack tactics, data-driven intelligence improves human experts' knowledge of cyber risks. This deeper understanding aids analysts in determining the seriousness of a threat and developing appropriate countermeasures specific to the organization's particular security posture. Ultimately, data-driven intelligence empowers cybersecurity analysts to support evidence-based, dynamic, and robust policy recommendations that strengthen an organization's resilience against cyber threats.

             \item \textit{Enhancing Human Experts' Domain Knowledge for Advanced Thinking:}
             Data-driven intelligence plays a pivotal role in enhancing cyber experts' domain knowledge, specifically for further modeling and analysis. By processing large volumes of cybersecurity data, data-driven tools can uncover valuable insights, patterns, and correlations that experts can use to build more accurate and sophisticated models. For instance, data insights can help determine which entities are essential for building an effective cybersecurity knowledge graph \cite{jia2018practical} or cybersecurity taxonomy building  \cite{mahaini2019building} through identifying common properties and characteristics of entities as well as their internal relationships. These data-driven models can capture the complexities of the cyber landscape, simulate various attack scenarios, and predict potential outcomes with higher precision. As cyber experts integrate data-driven intelligence into their domain knowledge, they can continuously refine their models, improve their understanding of cyber threats, and develop more effective strategies to defend against evolving challenges. Ultimately, the fusion of data-driven intelligence with the expertise of cyber experts enables them to create advanced models that are both robust and adaptable, empowering organizations to stay ahead in the ever-changing cybersecurity landscape.

             \item \textit{Knowledge Retention and Transfer:} Developing and maintaining efficient cybersecurity capabilities is a complex and continuing process that requires significant investment in terms of time, resources, and expertise. Professionals in the field of cybersecurity require not only technical skills but also a thorough awareness of the infrastructure, processes, and potential vulnerabilities within the organization. This knowledge is essential for quickly recognizing risks and taking appropriate action.
             In the real-world scenario, the expense of bringing cyber professionals is not only limited to salary and overheads but also the economic loss due to incidents which could have been better handled with experienced staff. Experience and such investments are lost momentarily when an experienced staff member leaves an organization. Consequently, this may result in a knowledge and expertise gap that is difficult to recover instantly. Numerous negative effects, such as increased vulnerability to cyberthreats, decreased efficacy of incident response, and potential project disruptions, may result from this loss. The hiring of a new professional with matching capabilities may not be sufficient because understanding the organizational context usually takes time and may result in further incidents. The data-driven approach creates new opportunities to retain this knowledge and experience and transfer them to new professionals within an organization as needed.
        \end{itemize}
   \end{enumerate}  

In summary, data-driven intelligence derived from raw data are crucial for both automating large-scale complex tasks and assisting human experts while making their decisions in the context of cybersecurity, illustrated in Figure \ref{fig: data-driven-intelligence}. It combines the strengths of data insights as well as AI and machine learning techniques for advanced modeling, highlighted in Section \ref{Data Insights and Modeling} to improve overall cyber defense capabilities and maximize teamwork between automated systems and human analysts. While each strategy has merits, a well-balanced approach that leverages both human expertise and data-driven automation to improve overall security posture and incident response capabilities could be the most effective way in cybersecurity. It enables human analysts to focus their attention on tasks that need critical thinking, creativity, and strategic planning by providing a wealth of data and insights.

\section{Data Insights and Modeling}
\label{Data Insights and Modeling}
This section mainly consists of two parts. We initially focus on different types of insights that are associated with data-driven intelligence, and then we concentrate on a general data-driven modeling workflow for further exploration to address a specific issue. 

\subsection{Cyber Data Insights}
For a better understanding of the insights involved in the data-driven intelligence process, we have highlighted three key questions in the section below. The answers to these queries can aid human analysts in deeper understanding and solving a specific problem in the context of cybersecurity as well as in automating the necessary tasks. These are:

\begin{itemize}
    \item \textit{What happened in the past?:} This typically explores the happenings and incidents in the world of cybersecurity. It includes analyzing historical data, logs, and incident reports to determine the type of cyberattacks, the methods employed by threat actors, the affected systems or data, and the overall impact on the organization. Experts in cybersecurity can react quickly, mitigate loss, and initiate the proper incident response procedures when they are aware of what happened. Building a strong defense strategy also involves identifying patterns and trends in cyber incidents.
    
    \item \textit{Why did it happen?:} Here, the emphasis is on underlying the root causes and associated factors for cybersecurity events. Understanding the ``why" requires an in-depth investigation of the security infrastructure's shortcomings, configuration issues, human errors, and vulnerabilities that led to the attack's success. Analysts can find systemic issues and weaknesses in their security procedures, work processes, and employee awareness using this investigative process. Organizations may improve their defenses, reduce potential risks, and build a more resilient cybersecurity framework by tackling these core causes.
    
    \item \textit{What will happen in the future?:} This element involves predicting and forecasting probable future cybersecurity threats and trends. Cyber threats are always changing, and threat actors are constantly coming up with new strategies to exploit vulnerabilities. Forecasting potential threats can be aided by data-driven intelligence, exchanging threat intelligence, and investigation of emerging technologies. Organizations can prepare for these challenges and be better able to protect themselves against new and emerging cyber threats by understanding what can happen in the future.
\end{itemize}

Thus extracting these insights could be the key to building the foundation of a data-driven intelligence model, where various techniques within the broad area of data science can be used discussed in the following.

\subsection{Data-Driven Modeling with Explanation}
An effective modeling technique is essential to extract insights or useful knowledge, where various data-preprocessing and visualization techniques as well as AI and machine learning algorithms for advanced modeling can be used. The key components of this process are as follows:

\begin{itemize}
    \item \textit{Data Collection and Preparation:} Gathering broad and comprehensive datasets related to cybersecurity is the first step. These datasets may contain information from various sources such as logs, network traffic, system events, security alerts, and historical attack data. To ensure consistency and quality, the collected data should be preprocessed, cleansed, and transformed towards the target solutions. Synthetic data generation as well as handling imbalanced issues using techniques like oversampling, and undersampling \cite{bagui2021resampling} might be helpful depending on the nature of the data.

    \item \textit{Feature Selection and Engineering:} This involves selecting or extracting meaningful features from the preprocessed data that can be used to build the model. It is essential to choose features carefully since traditional machine-learning methods, such as neural networks, SVMs, etc. are sensitive to the features used as inputs \cite{bakalos2019protecting}. The most pertinent features can be found through statistical analysis or machine learning algorithms \cite{sarker2021cyberlearning}. In some cases, feature extraction may require human expertise based on contextual information and awareness of cyber risks and vulnerabilities \cite{dick2019deep}. To identify relevant features and reduce the dimensionality of the data both algorithmic approach and domain experts may guide towards optimal feature engineering process. 

    \item \textit{Exploratory Analysis and Visualization:} Before moving on to advanced modeling or decision-making, this exploratory analysis helps in understanding in-depth data structure and patterns, and eventually to gain insights into normal behavior and identify patterns associated with cyber threats. Various statistical and visual techniques and tools such as Histograms, Scatter Plots, Bar charts, Heatmaps, etc. \cite{pedregosa2011scikit} can be employed to analyze the distributions, correlations, and structure of the data.

    \item \textit{Model Development and Training:} Models may vary depending on the characteristics of the data and fitting the problem domain. This includes applying AI and machine learning techniques like decision trees, random forests, neural network learning, as well as rule-based modeling and explanation \cite{lundberg2022experimental} \cite{sarker2021context}. To improve performance and generalization, optimizing model parameters is important. In several cases, innovative methods might need to develop based on what insights are needed to explore as discussed earlier. Developing hybrid or ensemble models that aggregate outcomes from multiple base models might need to take into account to improve model robustness and generalizability as well as overall accuracy.

    \item \textit{Model Evaluation:} A comprehensive evaluation is necessary after building and training the model with the relevant cyber data. The efficiency of the model can be assessed using evaluation criteria like accuracy, precision, recall, or F1 score \cite{sarker2021cyberlearning}. Validation methods like k-fold cross-validation aid in estimating the performance of the model on unseen data and evaluating its generalizability, which is important to take into account diverse real-world issues.

    \item \textit{Human-in-the-Loop Integration:} While automated models are capable of detecting a wide range of threats involved, they may not be flawless and could sometimes generate false positives or false negatives. Experts in cybersecurity may contribute their knowledge and expertise to the process by analyzing and verifying the outcomes of automated models. Thus, this module incorporates incident response teams and cybersecurity analysts in the process to provide domain expertise, interpret model outputs, and make critical decisions. 

    \item \textit{Deployment and Continuous Improvement:} The models can be deployed in a real-world cybersecurity context if they have been established to be satisfactory. To ensure the model's efficacy over time, it is essential to continuously assess its performance, detection rates, false positives, and false negatives. To keep the model realistic and up to date, regular updates, retraining, and adaptation to changing threats are required.
\end{itemize}

Overall, a comprehensive data-driven intelligence framework for cybersecurity modeling needs to be adaptable, resilient, and able to handle the constantly changing and evolving nature of cyber threats. To develop reliable and effective cybersecurity solutions, it thus needs to incorporate in-depth data analysis, machine learning, and domain expertise.

\section{Real-World Cybersecurity Application Areas}
\label{Real-World Application Areas}

Data-driven intelligence can be employed in various application areas for effective cybersecurity solutions. In the following, we summarize and discuss some important fields where data-driven intelligence could play a key role in both automation and assisting human experts in their decision-making process in various real-world applications.

\subsection{Critical Infrastructure}
Critical infrastructure (CI) typically refers to the systems, assets, and networks that are essential for the functioning of a society and economy, for example - energy, water, transportation, communications, healthcare, and finance are some of the potential sectors \cite{wisniewski2022industry} \cite{sectors-CI-Aus}. Thus, CI cybersecurity and resilience is one of the topmost important sectors nowadays, where data-driven intelligence could play a crucial role in practical solutions through data insights and sophisticated analytical modeling. The basis of intelligence could involve analyzing and visualizing CI data gathered from various sources including network logs, system activity, and threat intelligence feeds. The extracted insights from data could provide a comprehensive picture of the security landscape, providing human professionals with a better understanding of potential threats and vulnerabilities. Data-driven intelligence is also capable of predicting possible future cyber threats and attack trends using data patterns and AI algorithms \cite{sarker2022multi}. These predictive insights could be beneficial to human experts to further analyze the potential attacks and make countermeasures for them, enabling them to proactively strengthen CI defenses. In many cases, automaton is necessary because of speeding up the investigation and management of incidents as well as minimizing the possibility of human error. For example, routine incident response tasks, such as anomaly detection, malware analysis, and containment processes, could be automated through a data-driven modeling process. Overall, the potential of data-driven intelligence could be the key to next-generation CI security offering automating large-scale tasks as well as assisting CI professionals to make well-informed decisions in various real-world scenarios.

\subsection{Digital Twin}
Nowadays, more and more businesses are using digital twins, which are virtual replicas of physical assets or systems \cite{faleiro2022digital}. As physical, digital as well as communication space is associated with digital twin systems \cite{alcaraz2022digital}, an effective security measure is necessary. Data-driven intelligence may keep track of the network traffic, user interactions, and behavior of the digital twin \cite{kaur2020convergence}. Thus it enables real-time monitoring of digital twin systems, continuously collecting data to detect any deviations from normal behavior. Cyber professionals may gain deeper insights into the behavior of the physical and virtual components through this extensive data analysis. For instance, when any suspicious activity or possible security issues are identified, they may receive prompt notifications, enabling quick response and mitigation. It can also forecast potential cybersecurity risks and vulnerabilities based on the insights extracted from data. Overall, this could be a useful tool for automatically solving security issues as well as enhancing human expertise and aiding in their decision-making process in real-world applications.

\subsection{Smart Cities}
Smart cities could be another potential area, which typically rely on interconnected digital systems and devices to enhance efficiency and improve the quality of life for residents. Massive amounts of data are produced by smart cities from a variety of sources, including IoT devices, sensors, infrastructure, and human interactions \cite{sarker2022smart}. This data can be analyzed by data-driven intelligence to find trends and abnormalities that could point to possible cyber threats. It can identify suspicious activity in real-time and inform cybersecurity professionals, allowing them to take prompt action to stop cyberattacks. Data-driven intelligence may establish baseline behaviors for various parts of the smart city infrastructure by using AI and machine learning techniques \cite{sarker2022multi}. This involves being aware of the typical data exchange patterns, user behavior with regard to smart devices, and network traffic flow. Automated incident response systems may be triggered when cyber threats are identified. This can forecast potential future cyber threats through analysis of historical data and cyberattack trends. Decision-makers can comprehend how cybersecurity resources are used by conducting data analysis. Human experts could learn about emerging threats, observe trends, and make wise decisions about security practices and procedures according to this comprehensive picture.

\subsection{IoT}
The Internet of Things (IoT) enables communication and interaction with numerous devices, generates an enormous amount of data, which can then be utilized to identify trends, behaviors, make predictions, and conduct assessments \cite{hussain2020machine}. Thus decision-making in IoT cybersecurity is facilitated by data-driven intelligence, which substantially enhances human expert knowledge as well. Data-driven systems have the ability to rapidly detect abnormalities, recognize potential threats, and anticipate emerging issues by analyzing the enormous volumes of data produced by IoT devices and networks. This proactive strategy and real-time monitoring enable human professionals to react to cyber incidents quickly and strategically, reducing their effects. A thorough understanding of the complex IoT ecosystem is made possible by data-driven insights, which give important context and correlation from many data sources. This collaborative synergy enables cybersecurity experts to take well-informed decisions, allocate resources efficiently, and put into place efficient measures to protect IoT environments from emerging threats.

\subsection{ICS/OT}
ICS stands for ``Industrial Control Systems", and is typically used to monitor and control physical processes and operations, which typically connect IT components with sensors, actuators, and other operational technology (OT) devices \cite{conti2021survey}. Supervisory control and data acquisition (SCADA) systems, distributed control systems (DCS), PLCs, and other ICS components are frequently targets of cyberattacks \cite{bhamare2020cybersecurity}. Potential threats to the ICS include advanced persistent threats, supply chain compromise, distributed denial of services, etc., where data-driven intelligence can contribute to detect and mitigate through an extensive analysis. Utilizing real-time and historical data collected from numerous interconnected devices and networks within industrial infrastructure also enables human experts to gain an in-depth understanding of the evolving threat landscape. By analyzing patterns, anomalies, and potential vulnerabilities, experts may deal with cyber threats proactively before they escalate. Additionally, data-driven solutions enable routine and large-scale complex operations to be automated, allowing human experts stress-less. Overall, the security and reliability of crucial industrial systems could be ensured by developing effective defense modeling with the fusion of data insights and human expertise. 

\subsection{Metaverse}
Metaverse could be another potential area that can create secure, scalable, and realistic virtual worlds on a reliable and always-on platform. Users can interact with each other and digital objects in real-time using technologies like virtual reality (VR) or augmented reality (AR) \cite{huynh2023artificial}. Due to the massive volume of data moving around in the Metaverse, users are constantly running a higher risk of misuse \cite{wylde2023post}. Businesses are investing heavily in building an artificially intelligent Metaverse, which has increased the need for cybersecurity \cite{pooyandeh2022cybersecurity}. Data-driven cybersecurity solutions can track user behavior, interactions, and network traffic throughout the metaverse. These systems are capable of quickly identifying possible risks or unusual activities, such as unauthorized access attempts or malware activity, by analyzing patterns and anomalies. Automated incident response systems that can react to known threats and attacks without requiring human involvement could be provided by data-driven intelligence. Real-time monitoring and visualization of cybersecurity metrics and events within the metaverse can be provided through data-driven intelligence. These visualizations enable human professionals to promptly comprehend the security posture and pinpoint areas of concern. While human experts contribute critical thinking, domain knowledge, and decision-making, data-driven intelligence enhances these capabilities with rapid analysis, real-time insights, and automation of large-scale tasks. This can secure the metaverse environment in a comprehensive and proactive manner.

\subsection{Advanced Networking and Communications}
Nowadays, data-driven technology is also popular in the area of advanced communications and networking \cite{afzal2023data}. Based on current demand and traffic patterns, this can optimize the allocation of resources like bandwidth, computing power, and spectrum. In terms of security, data-driven technologies are capable of analyzing user behavior and network traffic patterns to detect anomalies and possible security breaches \cite{xu2019data}. Machine learning models can identify suspicious activity and trigger prompt countermeasures to stop intrusions. Predictive maintenance powered by data-driven intelligence enables proactive defense against evolving attack vectors, which can help prevent network downtime and improves overall reliability. Thus, this can ensure a balanced trade-off between security and usability, which dynamically adjusts security configurations based on network conditions, user behavior, and threat levels \cite{ahammed2023vision}. An effective access control system can be implemented by investigating user behavior and contextual data to ensure secure and reliable authentication. Overall, advanced communications and network security have been significantly impacted by data-driven intelligence as it provides insights, automation, and adaptation to address complex problems in real time.

In summary, organizations could enhance their threat detection and prevention, improve incident response capabilities, and strengthen their cybersecurity posture overall by utilizing data-driven intelligence in various real-world application domains. Data-driven intelligence augments human expertise rather than substituting it. In order to make informed decisions during security issues, it gives security analysts and operators more information and context. Data-driven intelligence helps human professionals to respond quickly and effectively by providing pertinent information and potential directions of action. Organizations can remain resilient in the face of emerging cyber threats when they have the capability to analyze massive datasets, uncover patterns, and make decisions based on data insights.

\section{Challenges and Research Direction}
\label{Challenges and Research Direction}
While the concept of data-driven intelligence revolutionizing the cybersecurity world holds promise, there are several challenges that researchers and practitioners need to address to fully realize its potential. These challenges discussed below encompass various aspects of research and development in the field:

\begin{itemize}
 \item \textit{Data Quality and Availability:} One of the major challenges in incorporating data-driven intelligence for cybersecurity research and applications is ensuring the quality and availability of relevant data. Obtaining comprehensive, accurate, and diverse datasets could be challenging, especially when dealing with sensitive information. Researchers need to overcome data limitations and address biases, as well as meaningful synthetic data generation to ensure the reliability and effectiveness of their research and ultimate outcome. Methods that enable cybersecurity models to transfer knowledge from one domain or task to another could be useful.

 \item \textit{Algorithmic Transparency and Interpretability:} The use of AI and complex machine learning algorithms, such as deep neural network learning \cite{sarker2022multi}, in data-driven intelligence may raise challenges in algorithmic transparency and interpretability. Understanding how algorithms make decisions and being able to interpret their outputs is crucial in the context of cybersecurity. Researchers need to focus on developing explainable AI techniques, e.g., rule-based modeling \cite{sarker2021context} or others that can provide insights into the reasoning behind algorithmic decisions, allowing cybersecurity professionals to trust and validate the results generated by data-driven intelligence systems.

 \item \textit{Privacy Concerns:} Data-driven intelligence might raise important privacy and ethical concerns. The collection and analysis of large amounts of personal and sensitive data need to be conducted responsibly and in compliance with privacy regulations. Researchers thus need to explore privacy-preserving techniques such as differential privacy, federated learning, data anonymization, etc. \cite{husnoo2021differential} to ensure that individuals' privacy is protected while still extracting meaningful insights from data. For instance, federated learning enables training models across numerous devices or organizations without sharing raw data, hence protecting data privacy.

 \item \textit{Adversarial Attacks and Defenses:} Adversaries can manipulate or poison datasets to mislead machine learning algorithms, leading to erroneous decisions or bypassing detection mechanisms. Research is necessary for developing robust models that are resilient to adversarial attacks and maintain high reliability and accuracy in practical settings. Developing advanced anomaly detection techniques identifying unusual behavior that can detect and respond to previously unknown and unseen threats and zero-day attacks is crucial. Hybrid models combining data-driven approaches such as machine learning, and rule-based approaches with expert knowledge can enhance the overall effectiveness of cybersecurity models.

 \item \textit{Generalizability and Scalability:} The effectiveness of data-driven intelligence models in cybersecurity may vary across different contexts, environments, and evolving cyber threats. Thus, ensuring the generalizability and adaptability of research findings and models is crucial. Investigating transfer learning techniques can assist models in maintaining high detection accuracy while adapting rapidly to new attack patterns. To manage huge datasets in real time, it is also necessary to develop scalable algorithms, distributed computing frameworks, and optimized processing strategies. This is crucial to assure scalability and efficiency due to the exponential growth of cybersecurity data volume.

 \item \textit{Human-in-the-Loop and Accountability:} While data-driven intelligence can provide valuable insights, the `human' element in real-world applications might not be overlooked. Researchers need to take into account how human operators interact with data-driven systems, understand their decision-making processes, and design effective user interfaces and visualizations to aid decision-making. Combining AI and human expertise can also increase accountability. For instance, cybersecurity professionals can validate model outcomes, intervene when necessary, and provide explanations for actions made by the AI system. Thus, a regulatory guiding framework comprised of data science researchers, cybersecurity experts, legal professionals, and policymakers is crucial to bridge the gap between technological breakthroughs and actual application.
\end{itemize}

In summary, data-driven intelligence for cybersecurity modeling is an area of study that involves resolving issues with data quality, processing techniques, model robustness, privacy, human expertise, and more. Addressing these challenges is crucial to fully realize the potential of data-driven intelligence in revolutionizing the cybersecurity landscape, which should be the key focus for future research and improvement.

\section{Conclusion}
\label{Conclusion}
This position paper has made a convincing argument for the revolutionary effects of data-driven intelligence in the cybersecurity area. For this, we have explored and discussed in-depth potential of data-driven intelligence, particularly, in terms of automating large-scale complex tasks as well as assisting human experts to make their decisions in real-world scenarios. Organizations can improve their capability to recognize and address emerging threats by utilizing the power of data intelligence. The proactive and adaptable nature of data-driven intelligence also allows security professionals to stay one step ahead of malicious actors, significantly reducing risks. However, this paradigm shift also includes several challenges such as data availability, algorithm bias, incorporating human expertise in the loop that are needed to be resolved, discussed in this paper. Building a well-balanced framework leveraging both human expertise and data-driven intelligence which can improve overall security posture, is also highlighted. Overall, we believe that data-driven intelligence could be the key to next-generation cybersecurity if it is deployed wisely and ongoing research is undertaken.

\section*{Acknowledgement}
This work is supported by Cyber Security Cooperative Research Centre (CSCRC), Australia.

\bibliographystyle{unsrt}
\bibliography{myref.bib}
\end{document}